\documentclass[twocolumn]{aastex631}
\usepackage{float}
\usepackage{subfigure}
\usepackage{silence}
\usepackage{amsmath}

\newcommand\species[2]{#1 {\sc #2}}

\begin{document}

\title{ High Lithium Abundance Connection with the Chromospheric Helium in Red Giants:  Spectroscopic and Asteroseismic analyses }

\correspondingauthor{Anohita Mallick, Bacham E. Reddy}
\email{anohitamallick@gmail.com,bachamr@gmail.com}

\author[0000-0002-4282-605X]{Anohita Mallick}
\affiliation{Indian Institute of Astrophysics, 560034, 100ft road Koramangala, Bangalore, India}
\affiliation{Pondicherry University, R. V. Nagara, Kala Pet, 605014, Puducherry, India}

\author[0000-0002-3456-5929]{Christopher Sneden}
\affiliation{Department of Astronomy and McDonald Observatory, The University of Texas, Austin, TX 78712, USA}

\author[0000-0001-9246-9743]{Bacham E. Reddy}
\affiliation{Indian Institute of Astrophysics, 560034, 100ft road Koramangala, Bangalore, India}

\affiliation{Department of Physics, Indian Institute of Technology Jammu, Jammu 181221, India}

\author[0000-0002-2516-1949]{Melike Afşar}
\affiliation{Department of Astronomy and Space Sciences, Ege University, 35100 Bornova, İzmir, Türkiye}

\begin{abstract}
We present a study of correlations between high Li abundances and strong chromospheric He I 10830 \r{A} absorption line strengths in $Kepler$ field giant stars. Our sample includes 84 giants with detectable solar-like oscillations in their lightcurves and their Li abundances come from the literature or measured here using LAMOST medium-resolution spectra. Evolutionary phases are determined through asteroseismic analysis, with mixed-mode period spacing ($\Delta$P) used to infer the time evolution of RC giants. Near-infrared observations of the He I $\lambda$10830 line were obtained with the high-resolution Habitable-zone Planet Finder (HPF) spectrograph on the Hobby–Eberly Telescope (HET). We find high Li abundances and strong He I lines exclusively among red clump (RC) giants, with their absence in red giant branch stars suggesting a shared origin linked to the He-flash. Additionally, a steady decline in He I strength with decreasing Li abundance among RC giants indicates a correlation between these properties. Older, Li-normal RC giants are He-weak, while most younger super-Li-rich giants are He-strong, suggesting temporal evolution of both phenomena. We hypothesize that the core He-flash and subsequent sub-flashes may enhance Li abundances in RC giant photospheres and trigger heightened chromospheric activity, leading to stronger He I $\lambda$10830 \r{A} lines in younger RCs. Over time, post-He-flash, chromospheric activity diminishes, resulting in weaker He I lines in older, Li-normal RCs.
\end{abstract}


\section{Introduction} \label{sec:intro}
Lithium (Li) is  one of the elements known to have primordial origin. Standard big bang nucleosynthesis (BBN) theories predict A(Li)~$\simeq$~ 2.72~dex\footnote{For elements X and Y, A(X)~$\equiv$ log~$\epsilon$(X) = log $(N_{X}/N_{H})$ + 12.0, and [X/Y] = log $(N_{X}/N_{Y})_{\star}$ -- log $(N_{X}/N_{Y})_{\sun}$. Metallicity will normally be assumed to be the [Fe/H] value.}, generally considered to be the primordial value \citep{2008Cyburt}. The measured high values of Li abundance of A(Li) $>$ 3.2 dex in very young stars or in the ISM suggests that the Galaxy has been enriched with additional Li since the big bang \citep{2009Asplund}. Cosmic ray spallation (CRS) and stellar nucleosynthesis are two of the major sources identified for Li enrichment in the Galaxy. CRS alone seems to be inadequate to explain the four-fold increase in Li \citep{1972mitler,2001Romano}. Moreover, canonical models do not predict Li production in stars \citep{1968Iben}.
In general stars are considered as Li sinks and the observations largely comply with the theory \citep{2000Pinsonneault}. Spectroscopic studies conducted over the past five decades have identified a small subset of evolved stars with exceptionally high lithium abundances. These include intermediate-mass asymptotic giant branch (AGB) \citep{1995Smith, 2020Holanda} and low-mass red giant branch (RGB) stars \citep{2011Kumar, 2011Alcala, 2021Martell}. The high Li in AGB stars is attributed to Hot Bottom Burning (HBB) \citep{1992Sackmann}. In contrast the origin of high Li in low mass red giants remains an unresolved puzzle since its discovery \citep{1982Wallerstein}.

Significant progress has been made in this area recently, driven by large-scale spectroscopic surveys. Studies have identified many Li-rich giants (LRGs) with Li abundances more than A(Li)~$\simeq$~1.5~dex, an upper limit set by standard theories for giants. There are now a few hundred LRGs and among those a few dozen are super Li-rich giants (SLRs) with abundance A(Li) $\geq$ 3.2~dex \citep{2011Kumar,2019Deepak,2019Singh,2021Magrini,2021Yan}. Following the suggestion of \cite{2011Kumar}  that Li production may be linked to the He-flash at the tip of RGB, studies focusing  on identifying the evolutionary phase of LRGs revealed that the majority of LRGs are red clump (RC) giants \citep{2019Casey, 2020Kumar}.  
Interestingly, all the SLRs for which evolutionary phases have been determined using asteroseismic analysis are found to be in the He-core burning phase \citep{2019Singh, 2021Singh}. Studies show 
a  strong circumstantial evidence that the high Li abundance among red clump giants may have originated during the short phase of He-flash \citep{2020Kumar, 2021Martell, 2021Singh, 2022Sneden, 2023Mallick}. 
\par
The physical mechanism of Li production and mixing processes during the He-flash phase are not well understood. Also, it is not clear whether the He-flash is the sole source of high Li among RC giants. There are few observations showing very high Li abundance among giants on the RGB, particularly among clusters \citep{2011Ruchti,2016Kirby,2021Magrini,2023Tsantaki}. 
If this is true one needs to understand whether there are multiple sites for Li production in red giants. It would be worth  determining the evolutionary phase of some of these RGB LRGs using asteroseismic data.

Here, we investigate whether the Li rich giants have any other unique observational characteristics. One possibility is He line strength, a concept first suggested by the serendipitous discovery of a strong chromospheric \species{He}{i} 10830~\AA\ absorption feature in a Li-rich (A(Li) $>$ 1.5 dex) red giant (see \citet{2021Sneden}). A subsequent survey by \cite{2022Sneden} found that $\sim$56\% of Li-rich 
field giants in their sample have similarly strong \species{He}{i} 10830\r{A} absorption features. This has opened a new avenue for uncovering further clues about the high Li abundances observed in a small fraction of red giants. In this paper we explore a further possible link : red giant evolutionary state from asteroseismological signatures. We present Li abundances, \species{He}{i} 10830 \r{A} line strengths, and 
asteroseismic parameters for 84 $Kepler$ Field giants. In \S\ref{sec:sample}, we discuss the stellar sample selection criteria. \S\ref{sec:LAMOSTobs} describes lithium abundance measurements obtained from LAMOST spectra, and \S\ref{sec:hpfobs} covers the acquisition and reduction of high-resolution \species{He}{i} 10830 \r{A} spectra. In \S\ref{sec:astero}, we investigated the asteroseismic properties of the selected red giants, while \S\ref{sec:HevsLi} examines the variations in helium and lithium among different evolutionary stages. Finally, \S\ref{sec:IRexcess_binarity} explores potential infrared excess, binarity and other chromospheric activity indicators.

\section{Sample Selection}\label{sec:sample}
\begin{figure}
\centering
\includegraphics[width=\columnwidth]{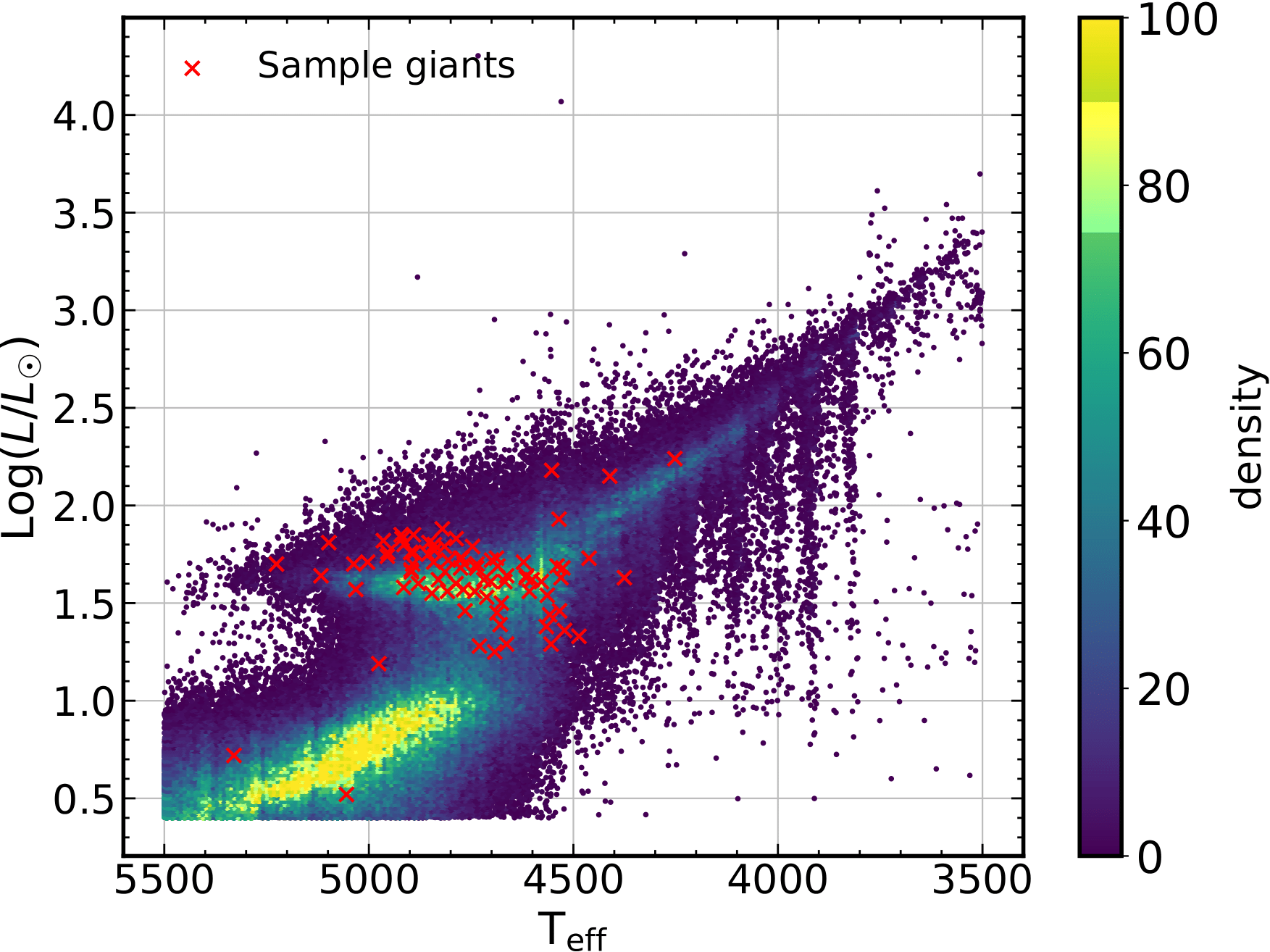}
\caption{HR diagram displaying the sample of 84 red giants (red crosses). Entire sample from $Kepler$ Input Catalog (KIC) is shown in the background. The colorbar represents the normalized star density in each region of the plot, with the maximum value scaled to 100. T$_{\rm eff}$ was taken from KIC \citep{2011Brown}, and luminosities were calculated using Gaia G-band magnitudes (see \citet{2018Andrae})}
\label{fig:sample}
\end{figure}

Our observational task was to gather high resolution spectra of \species{He}{i} 10830 \r{A} transitions in red giants that have asteroseismic data and either measured lithium abundances or spectra from which it can be derived. Following ground-breaking space-based asteroseismology efforts of the MOST \citep{2003Walker} and CoROT \citep{2001Catala} projects, the NASA $Kepler$ mission \citep{2010Borucki} observed more than half million stars, mostly centered on a single 115-degree field in Cygnus. In its final data release (DR25) \citep{2017Coughlin}, $Kepler$ detected 
solar-like oscillations in nearly 22,000 red giant stars \citep{2019Hon}.
\begin{figure*}
\gridline{\fig{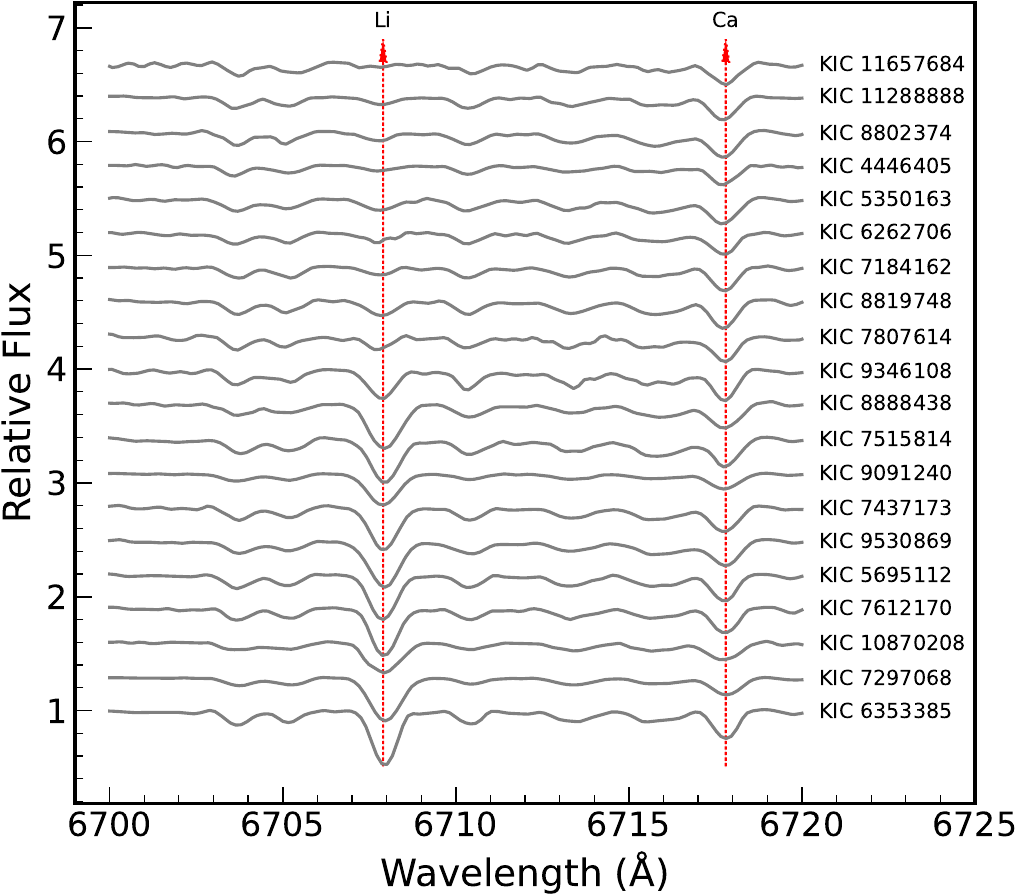}{\columnwidth}{(a)} \fig{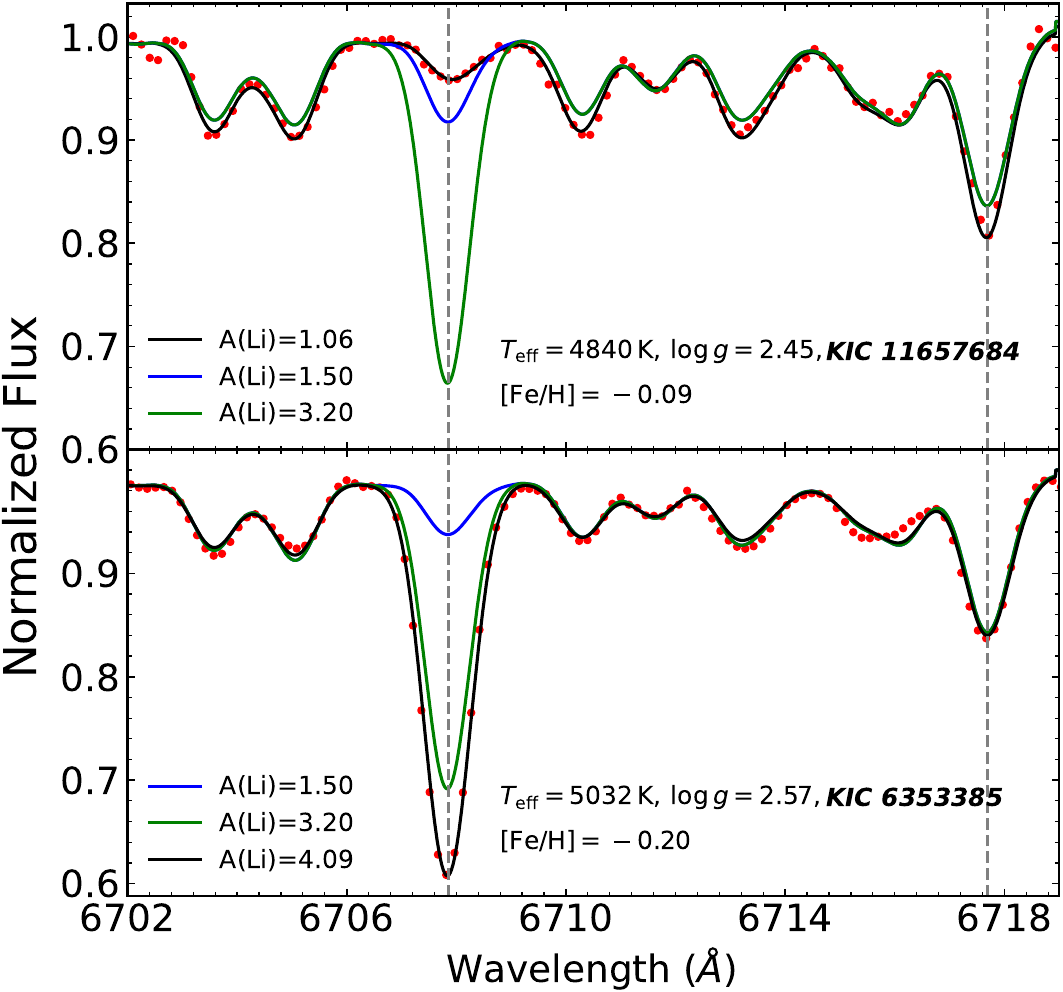}{\columnwidth}{(b)}}
\caption{In panel (a), spectra of a few giants from $LAMOST$ whose Li abundances were measured in this work.  Panel (b) illustrates spectrum synthesis for two sample $Kepler$ giants, representing the highest and lowest Li abundances measured. Observed spectra (red circles) are compared with the best-fit models (solid black lines) and additional models to illustrate the significance of Li detection: blue for A(Li) = 1.5 dex (classical Li-rich threshold) and green for A(Li) = 3.2 dex (SLR threshold).} The vertical dotted lines indicate the Li resonance line at 6707.8 \r{A} and a strong Ca line at 6717.7 \r{A}.
\label{fig:LiMRS}
\end{figure*}
We searched for $Kepler$ giants with published Li abundances, finding many in several recent studies \citep{2019Singh,2021Singh,2021Yan,
2017Takeda}. The lithium abundances reported in this work are derived from the LAMOST survey \citep{1996Wang}. The methodology employed for this analysis is outlined in Section \S\ref{sec:LAMOSTobs}.
From this list of $Kepler$ field red giants, we collected near infrared ($zyJ$ band, 8400$-$12500~\AA) high resolution spectra with the Habitable Zone Planet Finder Spectrograph (HPF) on the Hobby-Eberly Telescope (HET). 
The HR diagram of the sample, shown in Figure \ref{fig:sample}, highlights the distribution of these stars across the $Kepler$ field of view.\\
These spectra were employed to study their \species{He}{i} $\lambda$10830 \r{A} lines. We culled the sample to brightness range of 3~$<$~Jmag~$<$~13 so that we obtain IR spectra of optimal signal-to-noise ratio (SNR). In the end we collected HPF spectra for 84 stars (39 from the LAMOST survey). The complete sample for our study is provided in Table \ref{tab:1}.

\section{Lithium measurements from \texorpdfstring{$LAMOST$}{LAMOST}{}}\label{sec:LAMOSTobs}
We have extracted medium resolution spectra (MRS, R $\approx$ 7500) for 39 red giants from 
the LAMOST survey. Each MRS target provides a pair of spectra within a single exposure, consisting of blue (B) and red (R) band spectra spanning wavelength ranges of [4950 \r{A}, 5350 \r{A}] and [6300 \r{A}, 6800 \r{A}], respectively. We used the R-band spectra as they cover the \species{Li}{i} resonance line at 6707.8 \r{A}. Coadded spectra are available for all objects. 
All spectra have signal-to-noise ratios (SNR) in the R band exceeding 35, which is sufficient for abundance calculations. \par
The spectral data were brought to rest wavelength by correcting for stars radial velocity (RV) and continuum normalized using standard \texttt{IRAF} procedures.  Radial velocities are taken from $Gaia$ DR3 \citep{2023Katz}. The stellar parameters $T_{\mathrm{eff}}$, $log\ g$, [Fe/H] are extracted from the LAMOST MRS parameter catalog estimated by the LAMOST stellar parameter pipeline - LASP \citep{2015Xiang} . The microturbulent velocities ($\xi$) are estimated from empirical relations provided by \citet{2018Holtzman} and \citet{2016Garcia}. Utilizing these parameters, stellar atmospheric models were generated using the ATLAS9 code developed by \citet{2003Castelli}. Synthetic spectra were generated for each star based on their respective stellar parameters using the Python wrapper of the LTE radiative transfer code \texttt{MOOG} \citep{1973Sneden}, \texttt{pyMOOGi}\footnote{\url{https://github.com/madamow/pymoogi}}. Li abundances were adjusted in each spectrum to achieve the best fit with the observed spectra, minimizing the chi-square statistic. The resulting Li abundance (A(Li) was adopted as the final value for the program star. Figure \ref{fig:LiMRS} illustrates the spectra of selected giants analyzed in this study. Panel (a) shows representative spectra from LAMOST, highlighting the Li resonance line at 6707.8 \r{A} and the strong Ca line at 6717.7 \r{A}. Panel (b) contrasts spectrum synthesis for the stars with the lowest and highest Li abundances in our sample, demonstrating the spectral features used to determine A(Li). Derived values of A(Li) for all 39 stars have been provided in Table \ref{tab:1}.

\section{HPF observations and reductions}\label{sec:hpfobs}
We gathered high resolution HET/HPF spectra of 84 $Kepler$ giants.  The HPF is a near-IR spectrograph (zyJ photometric bands, 8100$-$12750~\AA).  Its development and working parameters have been presented in \cite{mahadevan12,mahadevan14}.\footnote{see \texttt{\url{https://hpf.psu.edu/}} for HPF parameter description}. HPF is an echelle spectrograph with 28 fixed spectral orders and resolving power $R$ $\equiv$ $\lambda/\Delta\lambda$~$\sim$~55,000.
Our spectra were obtained over a period of about two years.
\begin{figure}[!htbp]
\centering
\includegraphics[width=\columnwidth]{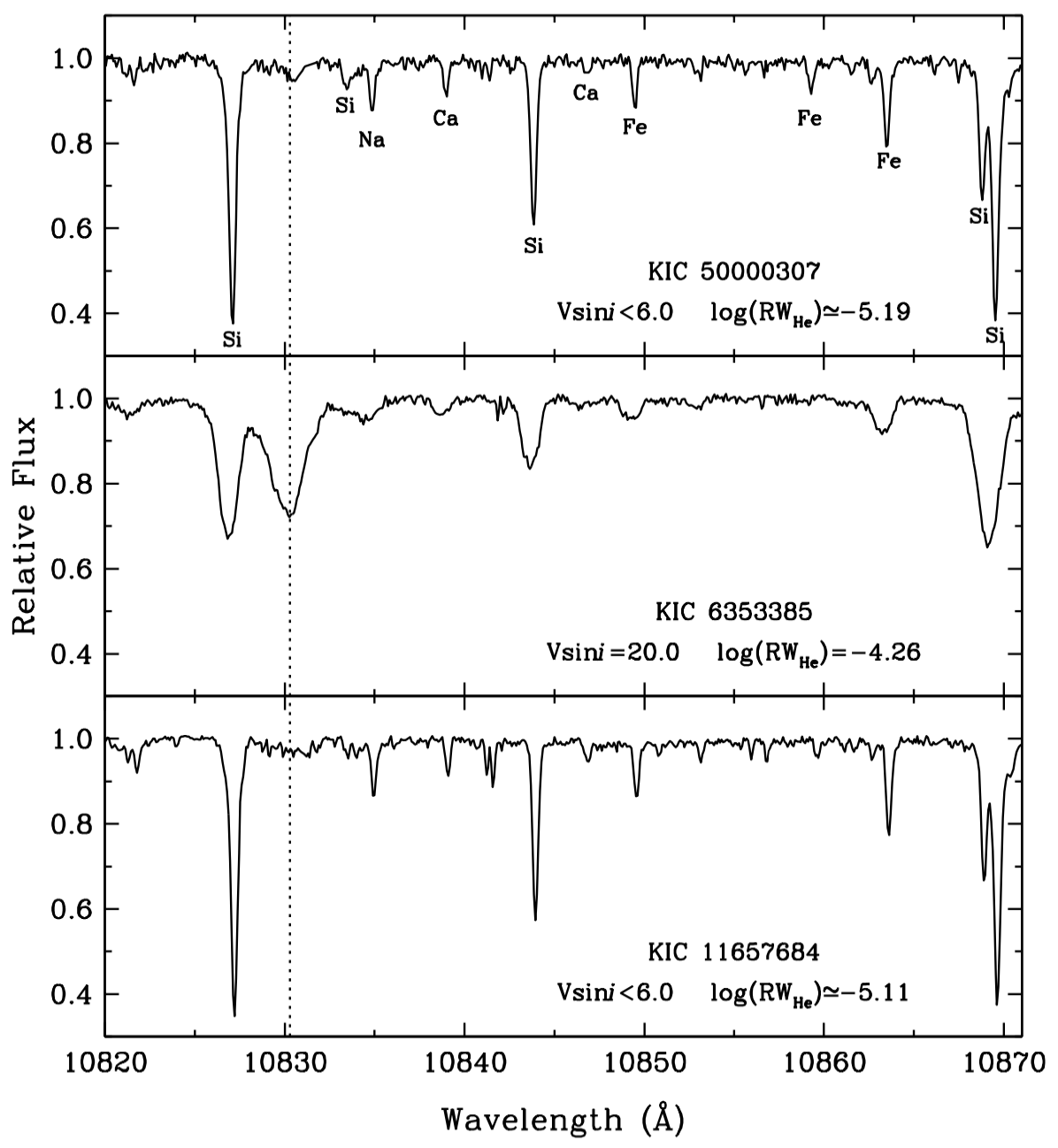}
\caption{\species{He}{i} 10830.3~\AA\ spectra of 3 program stars that appear in other figures of this paper.  The HPF spectral order containing the $\lambda$10830 line extends from about 10820~\AA\ to 10960~\AA, leading to the appearance of $\lambda$10830 near the blue end of the order.  Some prominent atomic features, all due to neutral species transitions, are labeled by element name, while the \species{He}{1} feature is indicated by a dotted vertical line.}
\label{fig:kep3spec}
\end{figure}
The HPF facility reduction package $Goldilocks$.\footnote{https://github.com/grzeimann/Goldilocks\_Documentation} operated automatically on the raw data frames to produce output files ready for reduction steps. We used IRAF \citep{tody86,tody93}\footnote{https://iraf-community.github.io/} routines to accomplish all steps leading to final 2D echelle spectra, including sky emission line subtraction, order-by-order continuum normalization, telluric absorption line division, wavelength scale transformation, and correction to rest velocity. In Figure~\ref{fig:kep3spec} we show example spectra of 3 of our program stars.\par

Analysis of the reduced spectra was limited to estimation of rotational velocity and equivalent width of the $\lambda$10830 line.  To derive these quantities we followed the methods discussed in detail by \citet{2022Sneden} (see further discussion by Af{\c s}ar, in preparation).  To summarize the procedure briefly, the first step was recognition that the \species{He}{i} $\lambda$10830 transition arises in red giant chromospheres, not photospheres.  This is due both to the 19.8~eV excitation energy of its lower state, and lack of connection to the ground state $-$ it is a metastable level (e.g, see Figure~3 of \citealt{preston22}).  But as illustrated by the spectrum of KIC 6353385 in Figure~\ref{fig:kep3spec}, strong $\lambda$10830 chromospheric lines have significant spectral overlap with nearby photospheric lines, especially \species{Si}{i} 10827.1~\AA.
The procedure involved creating synthetic spectra to model and remove contaminating photospheric lines near the \species{He}{i} $\lambda$10830 feature. The equivalent width (EW) was then determined by comparing the observed spectra with the synthetic ones, accounting for various broadening effects, including rotational, instrumental, and macroturbulent. For stars with detectable rotation, additional rotational smoothening was applied to the synthetic spectra, and the broadening parameters were adjusted iteratively to achieve the best match.

\section{\texorpdfstring{Asteroseismic investigation of $Kepler$ Red Giants}{Asteroseismic investigation of Kepler Red Giants}\label{sec:astero}}
\subsection{Stellar pulsation theory}

Solar-like oscillations occur in cool stars with outer convective envelopes. Turbulent motions in their convective zones trigger envelope pulsations deforming the surface. At the end of the RGB phase, stars with masses $\gtrsim$ 0.8 M$_{\odot}$ undergo the He-flash, leading to a rapid contraction in size and a decrease in luminosity. Post He-flash, stars settle into the core helium-burning phase known as the red clump (RC) \citep{1968Iben} or red horizontal branch (RHB). 
These stars occupy a very narrow luminosity range and exhibit slight variations in temperature due to differences in stellar mass and composition. The RC stars in a $T_{\rm eff} - L $ plot overlap with the giants ascending the RGB (see Figure \ref{fig:sample}), making it challenging to distinguish between them, 
especially in field stars. Asteroseismic analysis \citep{2011Bedding} has made it possible to accurately distinguish between RGB stars and RC giants. Two key asteroseismic parameters $-$ the p-mode large frequency separation ($\Delta\nu$) and the average period spacing ($\Delta P$) of dipole mixed oscillation modes $-$ can be used to differentiate these two stellar populations. The RC giants generally show higher $\Delta P$ than RGBs.
\begin{figure}[H]
\centering
\includegraphics[width=\columnwidth]{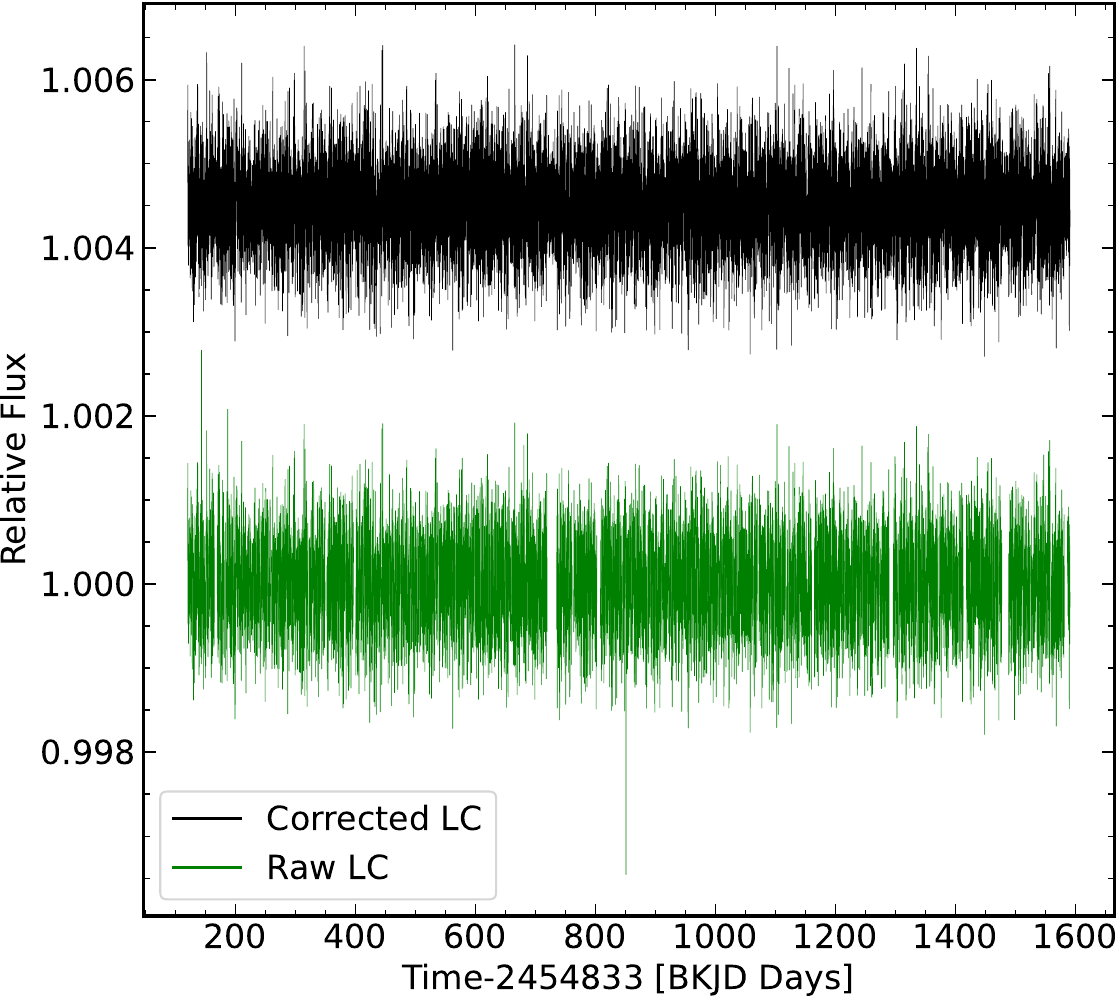}
\caption{The raw (green) and corrected (black) stitched lightcurves from $Kepler$ Q0-Q17 quarters for KIC 5000307. The corrected light curve has been vertically offset by 4.5 $\times$ 10$^{-2}$ for comparison
\label{fig:KIC5000307_data}}
\end{figure}
\subsection{Data preparation}
$Kepler$ space telescope observations consist of a range of pulsating stars with photometric variations monitored at two cadences - the short cadence (SC) of 58.9 sec observations or the long cadence (LC) of $\sim$ 29.4 minutes. Evolved RGB stars exhibit $\nu_{\text{max}} \sim$ 20 $\mu$Hz (the frequency at which oscillation modes reach maximum power), equivalent to half-day periods, making the 30-minute sampling rate sufficient. In this work, we have used LC lightcurves as the long-duration data are useful for detecting low-frequency oscillations and have better mode resolution. For all stars, $Kepler$ provides two types of fluxes : the raw pixel data, which is calibrated and photometrically analyzed - the Simple Aperture Photometry (SAP) flux with instrumental jitters, and the flux that has been systematically corrected for instrumental perturbations - the Pre-search Data Conditioning Simple Aperture Photometry (PDCSAP) flux \citep{2012Smith}. Three of our stars do not have $Kepler$ time series data for which we obtained $\sim$ 30min cadence data from $TESS$. $Kepler$ and $TESS$ lightcurves were processed using the \citet{2018Lightkurve}\footnote{\url{https://lightkurve.github.io/lightkurve/index.html}} package.

Although a quality masking process filters out most bad data points in the time series within the PDCSAP flux, certain issues can persistently affect the light curves. These include fluctuations in flux caused by cosmic rays, zero crossing events, Argabrightening from detector saturation \citep{2009VanCleve}, deviations due to the loss of fine pointing, and anomalies attributed to rolling band artefacts from detector electronics. A stringent 4.5 sigma clipping technique was applied to remove outlier data points caused during momentum desaturation \citep{2014Handberg}. Subsequently, only data points with quality flags set to zero were retained. Random white Gaussian noise was introduced to address any resulting data gaps. All the corrected lightcurves from different quarters were normalised and finally stitched together, which are suitable for asteroseismic analysis. We present raw and corrected stitched lightcurves for KIC 5000307 in Figure \ref{fig:KIC5000307_data} to illustrate the data preparation process critical for asteroseismic analysis.
\subsection{Detection of seismic parameters}
\begin{figure*}
\subfigure[]{\label{fig:4a}\includegraphics[width=0.5\textwidth,height=9cm]{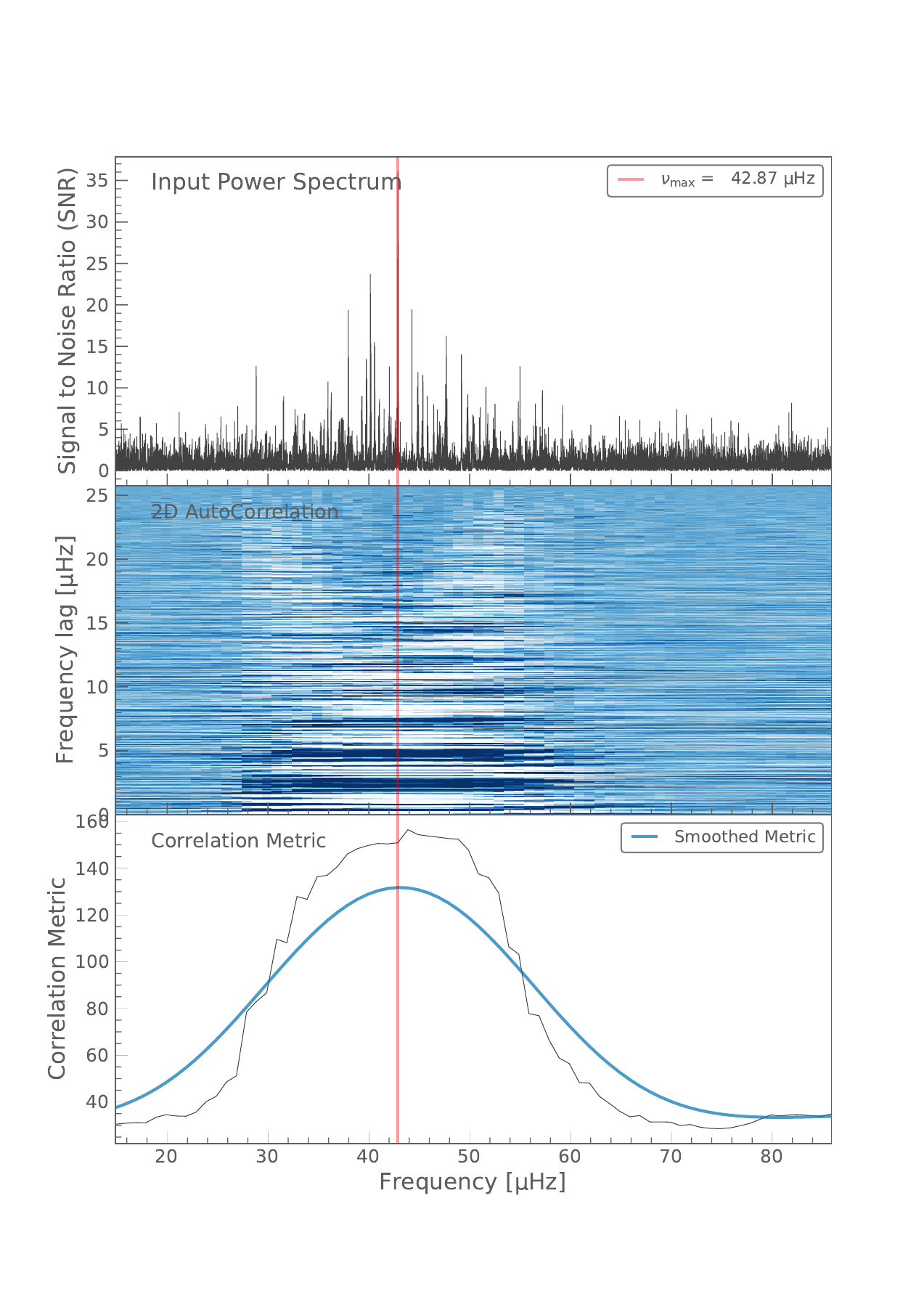}}
\subfigure[]{\label{fig:4b}\includegraphics[width=0.5\textwidth,height=9cm]{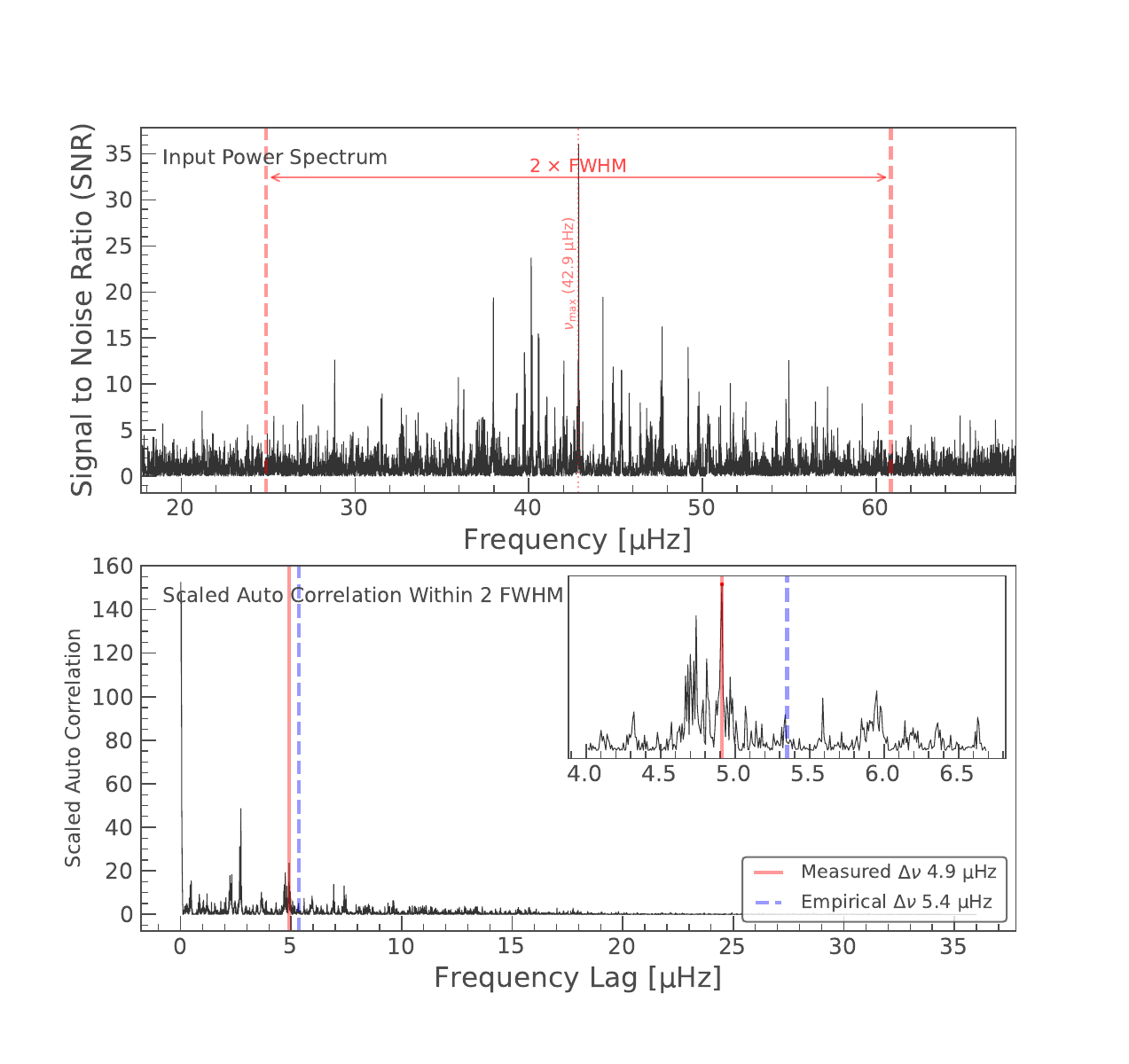}}
\caption{Results of Lightkurve analysis for KIC 5000307. In panel (a) estimation of $\nu_{max}$ using a smoothed 2D ACF over background corrected PSD. In panel (b) Peaks in ACF in the region near empirical $\Delta\nu$ for calculating $\Delta\nu$}
\label{fig:KIC5000307_lightkurve}
\end{figure*}

In asteroseismology, time series data are analysed in the frequency domain by calculating the power spectral density (PSD). To account for irregularly sampled lightcurves, the Lomb-Scargle periodogram technique is employed to estimate the PSD \citep{1976Lomb, 1982Scargle}. The PSD shows the signal amplitude over a range of frequencies. 
To estimate $\nu_{max}$, a small region in the background noise corrected PSD showing strong power excess is selected. 
The central peak frequency of this distribution is denoted as $\nu_{max}$. An empirical relation proposed by \citet{2009Stello} provides a rough approximation for $\Delta\nu$ :
\begin{equation*}
\Delta\nu_{est} = (0.263 \pm 0.009) \nu_{\text{max}}^{(0.772 \pm 0.005)} \, \mu\text{Hz}
\end{equation*}
A 2D auto correlation function (ACF) is computed within the same region which cross-correlates the data with a temporally shifted version of itself. 
As shown in Figure \ref{fig:4a}, the smoothed 2D ACF derived from the power spectral density (PSD) was instrumental in identifying $\nu_{max}$ for the example star KIC 5000307. Figure \ref{fig:4b} shows the ACF peaks near the empirical $\Delta \nu$ estimates for the same.\par
\begin{figure*}[!htbp]
\centering
\includegraphics[width=\textwidth]{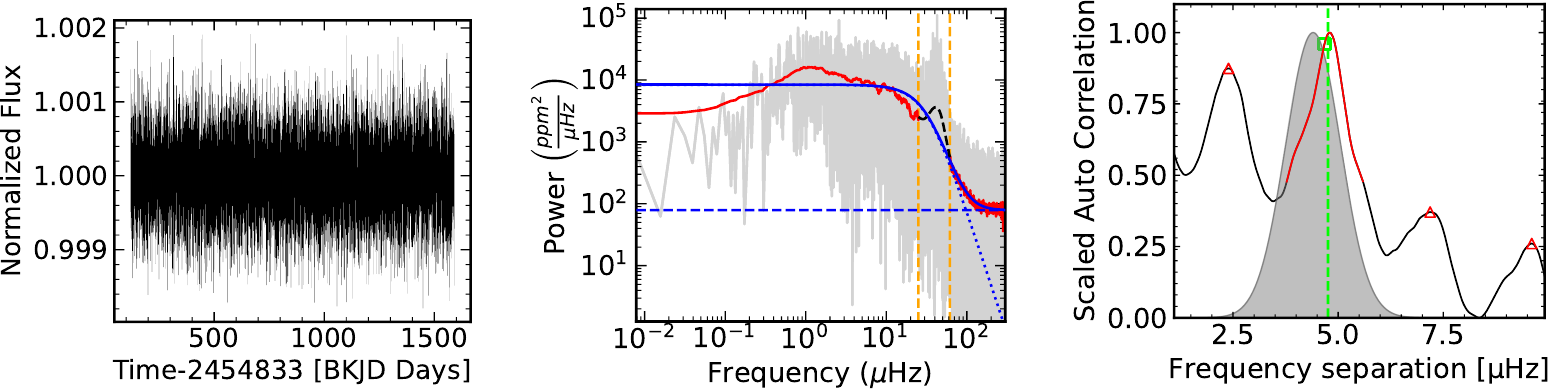}
\caption{\small pySYD results for KIC 5000307. 1st panel shows the corrected lightcurve, 2nd panel is the PSD where original PSD is shown in gray, the red curve is the smoothed PSD using a boxcar filter of 5 $\mu$Hz, black dashed line indicates the Gaussian power-excess superposed on the smoothed PSD. The blue dashed line indicates the white noise, blue dotted line shows stellar granulation and solid blue line is the overall best fit to the background. In the last panel is an ACF of a small window of the background corrected PSD centered on $\nu_{max}$. Black solid line is smoothed background-corrected PSD, red region indicates the extracted ACF peak. Gray shading represents the Gaussian weighting function to define the red region and the center of the Gaussian fit (green dashed line) provides the estimated value of $\Delta\nu$.}
\label{fig:KIC5000307pysyd}
\end{figure*}
When analyzing giant stars with low $\nu_{max}$ values, a frequency window width narrower than $\nu_{max}$ should be chosen to prevent over-smoothing of the PSD. However, 
\texttt{Lightkurve} cannot accurately fit Gaussians in narrow ranges which in turn affects the computation of ACF for stars with low $\nu_{max}$. Additionally, it does not support estimating the uncertainties of $\nu_{max}/\Delta\nu$. To address these challenges as well as for a recheck on our parameter estimates, we reanalyzed our entire sample with \texttt{pySYD}\footnote{\url{ https://github.com/ashleychontos/pySYD}}, an open source Python translation of the widely tested IDL based SYD pipeline \citep{2009Huber} developed by \citet{2022Chontos}. The primary difference between pySYD and \texttt{Lightkurve} is the modeling of background noise. \texttt{pySYD} employs Harvey-like functions along with white noise to fit the background due to stellar granulation activity. Power spectra of KIC 5000307 with the best-fit background (solid blue line) is illustrated in Figure \ref{fig:KIC5000307pysyd}, 2nd panel. The process iteratively models the best background fit which minimizes the Bayesian Information Criterion (BIC). Subsequently the methods for estimating $\nu_{max}$ and $\Delta\nu$ from the background corrected PSD remain consistent with \texttt{Lightkurve}. For calculating uncertainties a  Monte Carlo sampling introduces stochastic noise to the PSD. The background is iteratively fitted to the perturbed PSD and global seismic parameters are recomputed for $\sim$ 200 times. 
Figure \ref{fig:KIC5000307pysyd} displays the \texttt{pySYD} analysis results for KIC 5000307, including the corrected lightcurve, background-corrected PSD, and derived seismic parameters, providing more robust background corrections and uncertainties compared to the results from \texttt{Lightkurve} shown in Figure \ref{fig:KIC5000307_lightkurve}.

\begin{figure}
\centering
\includegraphics[width=\columnwidth]{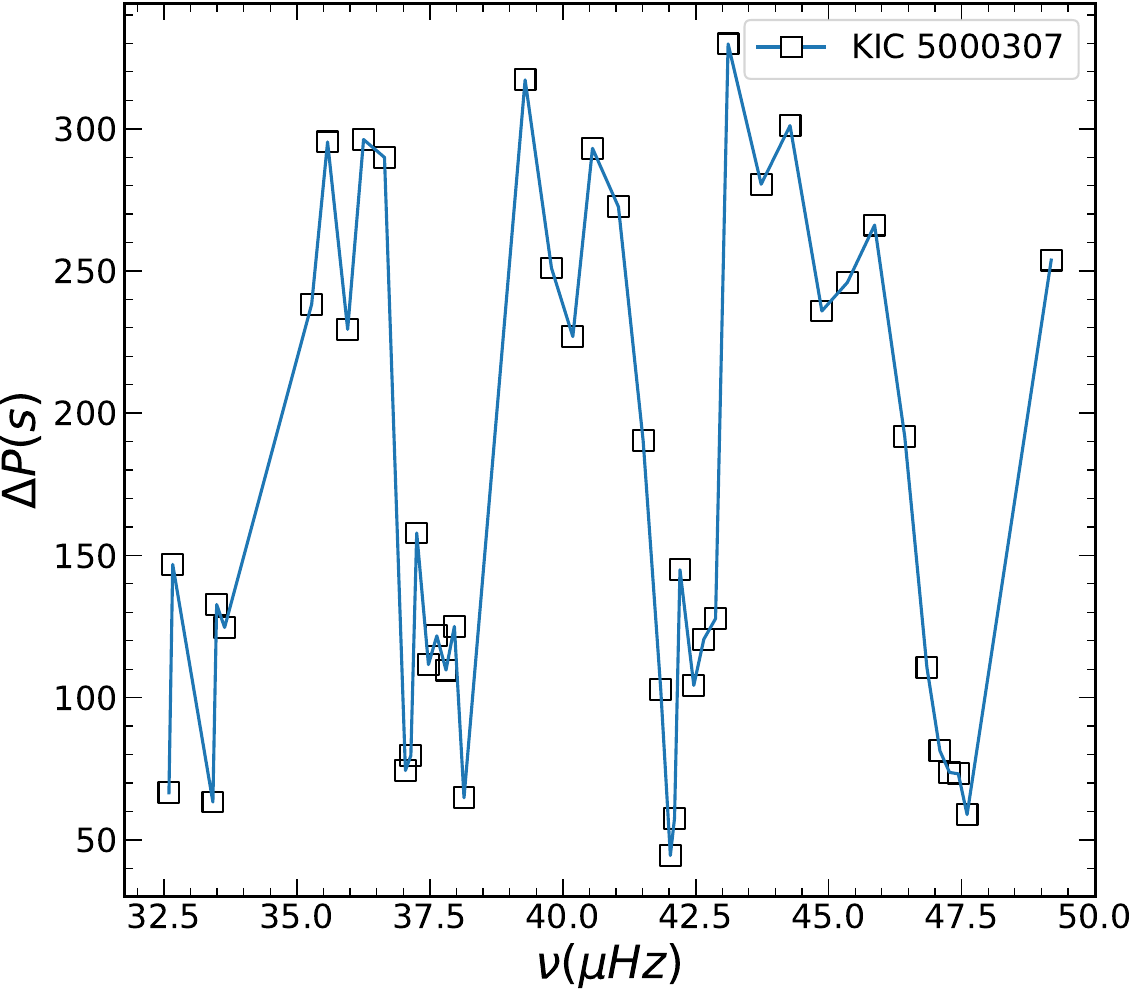}
\caption{Period spacing for KIC 5000307}
\label{fig:KIC5000307_period}
\end{figure}
For estimating $\Delta$P, the background corrected PSD is again smoothed using a Gaussian filter ($\sigma$ $\sim$ 2). An initial guess of mixed dipole mode ($l$=1) frequencies is made by identifying peaks in the smoothed flux data by comparing values to their neighbors. Regions are selected containing at least 4-5 consecutive $l$ =1 modes manually. The periods between consecutive $l$ =1 frequencies is computed. The average and standard error of these periods is propagated as the average mixed mode period spacing $\Delta$P and its uncertainty as illustrated in Figure \ref{fig:KIC5000307_period}. Seismic parameters for all stars are shown in Table \ref{tab:1}.

\subsection{Evolutionary Status}
\begin{figure}
\centering
\includegraphics[width=\columnwidth]{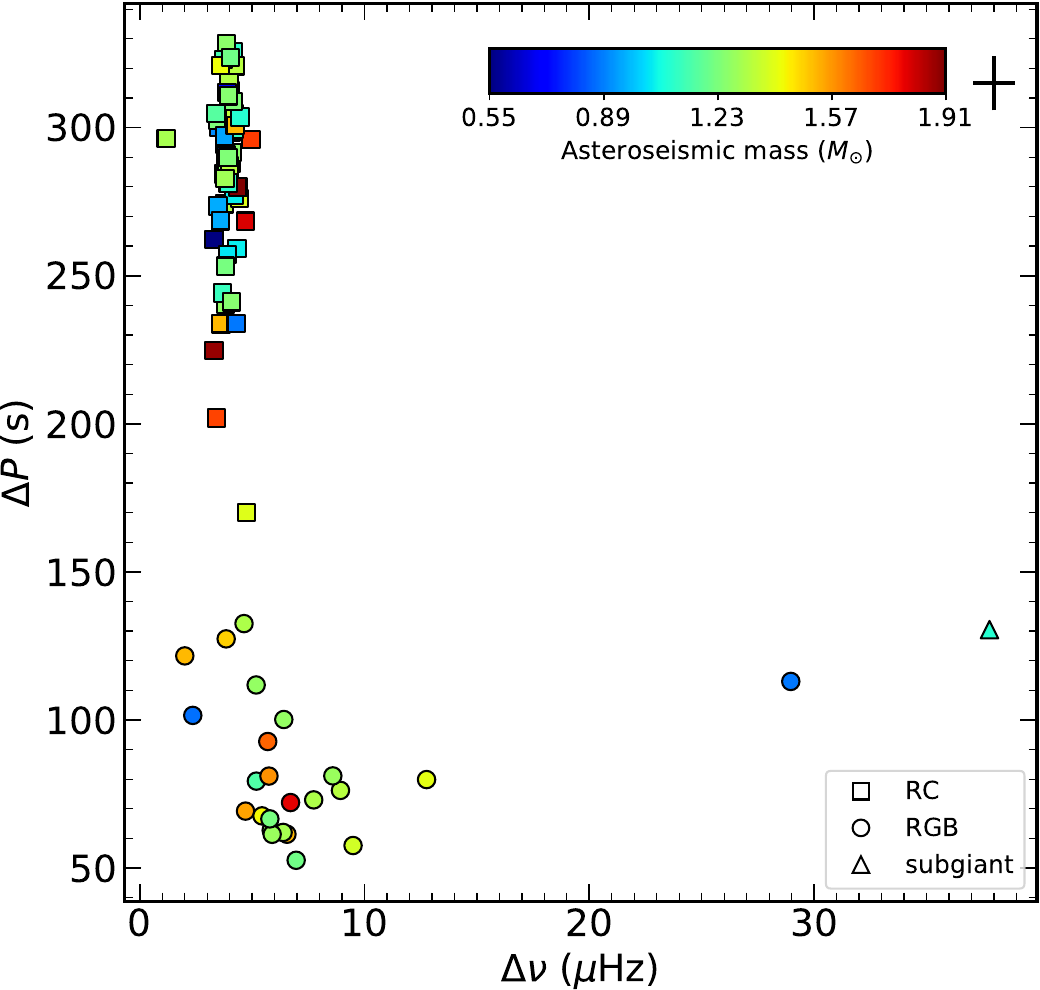}
\caption{Average $\Delta$P vs $\Delta\nu$ for all stars. The error cross at top right indicates typical uncertainties in $\Delta$P and $\Delta\nu$}
\label{fig:Pvsnu}
\end{figure}

In the $\Delta$P-$\Delta\nu$ diagram (Figure \ref{fig:Pvsnu}), RGB stars occupy the lower $\Delta$P regime. Following the classification criteria by various works  \citep{2016Vrard,2018Ting}, we adopted all stars with $\Delta P \ <$150s  as red giants in the H-burning phase and stars with $\Delta$P $\geq$ 150s as RC giants in the Core He-Burning (CHeB) phase. Among the H-burning stars, we have 24 RGB and 1 subgiant star \citep{2014Mosser}. In total the sample has 59 CHeB, 24 RGB and 1 subgiant. We calculated seismic stellar masses using the corrected scaling relations given by \citet{2016Sharma}
\begin{equation*}
\frac{M}{M_\odot} \approx \left(\frac{\nu_{\text{max}}}{f_{\nu_{max}}\nu_{\text{max},\odot}}\right)^3 \left(\frac{\Delta\nu}{f_{\Delta\nu}\Delta\nu_\odot}\right)^{-4} \left(\frac{T_{\text{eff}}}{T_{\text{eff},\odot}}\right)^{3/2}   
\end{equation*}
\section{Helium and Lithium variations among RGB and CHeB stars}\label{sec:HevsLi}
\begin{figure*}
\centering
\includegraphics[width=\textwidth]{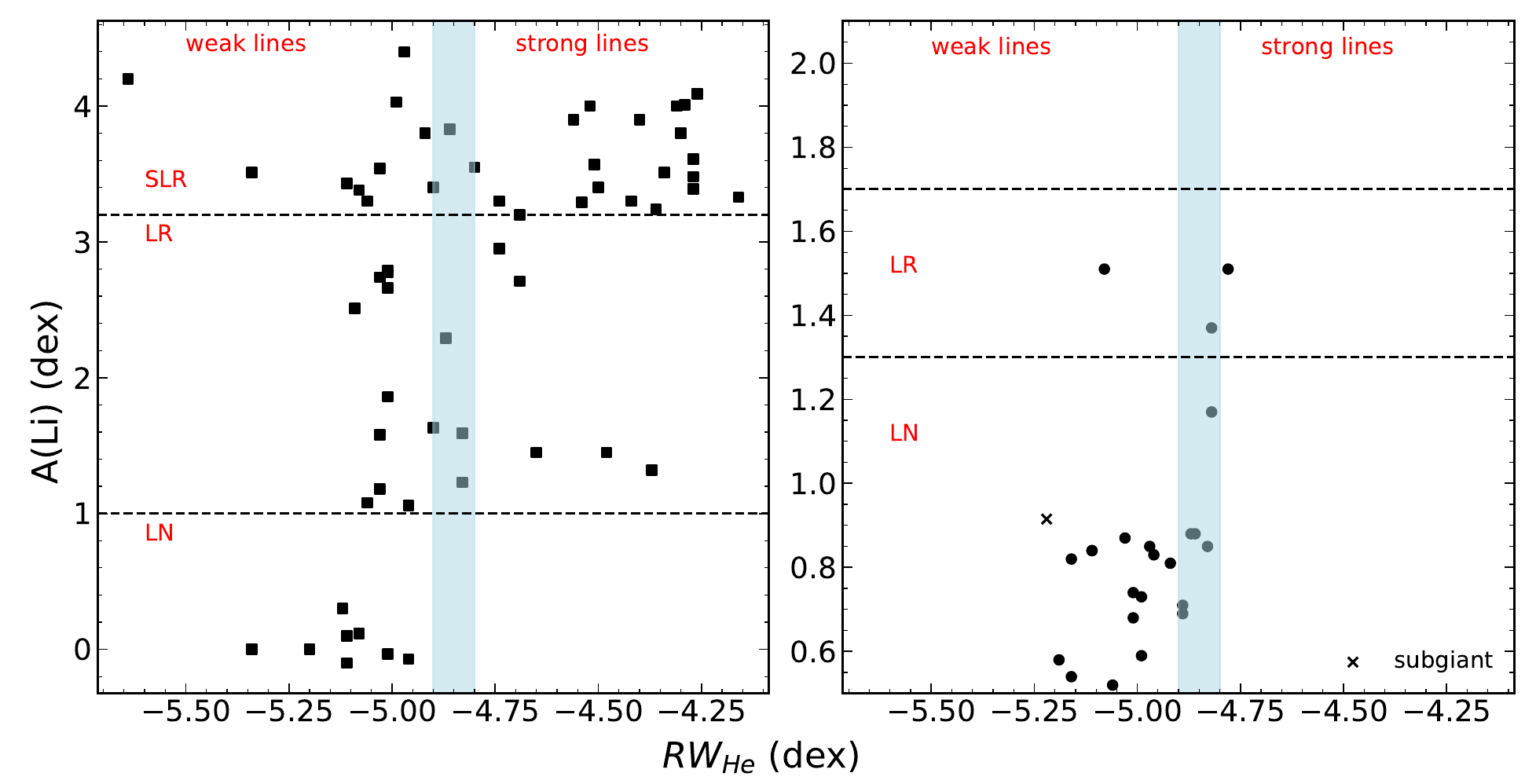}
\caption{Correlation between $RW_{He}$ and Li abundances. Panel on the left showcases RC stars, while the panel on the right showcases RGB stars. Both plots feature a vertical blue-shaded region that distinguishes between weak and strong $\lambda$10830 absorption strengths. Additionally, black dotted lines are used to categorize the stars into Li-normal, Li-rich, and SLR groups. Notably, the right plot identifies a single subgiant among the RGBs, marked with a $\times$.}
\label{fig:LivsHe}
\end{figure*}
EWs of the \species{He}{i} $\lambda$10830 lines for all stars were determined 
using a spectrum analysis software package \texttt{SPECTRE}\footnote{\url{http://www.as.utexas.edu/~chris/spectre.html}} \citep{1987Fitzpatrick}. Since there is a significant variation in the measured EWs of the sample giants (25-750 m\r{A}), we adopted logarithmic reduced widths:
$$RW_{He}=\log_{10}\left(\frac{EW_{He}}{\lambda}\right)$$ where EW is in \r{A}. The value $RW_{He}$ = -4.85 was adopted as the threshold for classifying stars with either weak or strong $\lambda$10830 transitions \citep{2022Sneden}. To understand the relation between chromospheric \species{He}{i} strength and photospheric Li abundance among RGB and CHeB stars, we grouped the sample stars based on the amount of Lithium in them. Since the definition of Li-richness varies with the evolutionary stage of the stars, a stage-specific classification is required \citep{2016Kirby}. Following the study by \cite{2021Singh}, we divided RC stars into three groups: Li-normal (RC$_{\rm LN}$) (A(Li) $\leq$ 1.0), Li-rich (RC$_{\rm LR}$)  ( 1 $<$ A(Li) $<$ 3.2 ) and
Super Li-rich (SLR) ( A(Li) $\geq$ 3.2~dex). In the case of RGB stars, we made them into two groups: Li-normal (RGB$_{\rm LN}$) ( A(Li) $\leq$ 1.7~dex) and Li-rich (RGB$_{\rm LR}$) (A(Li) $>$ 1.7~dex) \citet{2014Liu}. 
In Figure \ref{fig:LivsHe}  we showed the entire sample in a plot of A(Li) and $RW_{He}$ for both RC and RGB stars. Of all 58 CHeB stars, 31 exhibit weaker He lines, while 27 display stronger He lines. However, within the 31 SLR group stars, majority (20) show  stronger \species{He}{i} profiles compared to 11 stars with weaker \species{He}{i} line. Among 19 RC$_{\rm LR}$ stars,  12 have weaker He lines, while 7 display stronger lines. The subgiant star shows weak He absorption. In RC$_{\rm LN}$ group, all members exhibit weaker He lines. Among the 24 RGB stars, 3 are situated at the vertical shaded region, while the rest 21 stars exhibit weak signal.
\section{\texorpdfstring{Lithium and \species{He}{i} $\lambda$10830 Strengths: Influencing Variables}{Li and He I lambda 10830 Strengths: Influencing Variables}}
\label{sec:IRexcess_binarity}
In addition to the He-flash, there are alternative proposals for elevated Li among RC giants such as mergers or binary interactions. Several recent works offer insights into the mechanisms driving Li enhancement during binary evolution. \citet{2019Casey} proposed that tidal spin-up from a binary companion could instigate internal mixing, thereby triggering lithium production through \citet{1971Cameron} mechanism. According to this model, such enhancement might occur randomly at any point along the RGB or the clump phase. Alternatively, \citet{2020Zhang} suggest that Li-rich stars in the CHeB phase could be produced through mergers in a RGB+ Helium White Dwarf binary system, where the transfer of angular momentum from a companion leads to the ejection of stellar material, increased stellar rotation, and the formation of dust grains that result in infrared excess. Further observational support comes from \citet{2024Singh}, who found that a star exhibiting high A(Li) and rotational velocity, along with evidences of a binary companion, likely underwent tidal synchronization following the He-flash. Likewise, \citet{2024Susmitha} studied metal-poor super Li-rich giants and proposed that past mergers, rather than binarity alone, could explain elevated lithium levels, especially during the core helium-burning phase, and suggested that these stars might be in the early asymptotic giant branch phase. \par
We tried to investigate the presence (or absence) of binary companions and infrared excess in our sample by checking for photometric and astrometric variations. 
\subsection{IR excess}
We gathered optical, near-infrared (NIR), and mid-infrared (MIR) photometry data from the VOSA filter repository (Virtual Observatory SED Analyzer), developed by the Spanish Virtual Observatory project. These data were employed to construct optical-IR spectral energy distributions (SEDs) for all sources. We determined the photospheric contribution to the SEDs using \citet{2003Castelli} model atmospheres. Stellar parameters were either derived from existing literature or obtained from LASP (for stars with calculated lithium abundances in this study). VOSA determines the slope of the linear regression of the stellar SED iteratively, adding new infrared data points. It flags an object for IR excess if the slope is significantly smaller ($<$ 2.56) than expected from stellar photospheric emission. VOSA further refines IR excess by comparing observational and synthetic flux at each photometric point, identifying significant ($>$ 3$\sigma$) deviations as indications of IR excess. A more extensive explanation is available in the VOSA documentation.\footnote{\url{http://svo2.cab.inta-csic.es/theory/vosa/helpw4.php?otype=star&action=help&what=&seeall=1}} None of the sample giants show any NIR/MIR excess. However, given the absence of far-infrared data for our objects, we cannot dismiss the possibility of cooler dust around them.

\subsection{Binarity}
We examined our sample for the presence of astrometric and eclipsing binaries. The GAIA Renormalized Unit Weight Error (RUWE) serves as a valuable indicator for identifying astrometric binaries, with a threshold value exceeding 1.4 suggesting the presence of unresolved binary systems \citep{2023Halbwachs}. Only two stars KIC 10716853 and KIC 10404994 have RUWE $>$ 1.4. We also crossmatched our sample with the Hipparcos–Gaia Catalog of Accelerations (HGCA) \citep{2021Brandt} to leverage their longer baseline  for identifying binaries with slightly larger separations. Only four stars from our sample are present in the HGCA. HGCA provides $\chi^2$ values for a constant proper motion model with 2 degrees of freedom. We converted this $\chi^2$ value to a format similar to Gaia RUWE. Following \citet{2022Sneden}, HGCA RUWE values $>$ 3 suggest the presence of long-term astrometric variations. The same two stars also display HGCA RUWE $>$ 3. However, both the stars are in the RC phase with  weak He lines and low Li abundances.\par
To identify potential eclipsing binaries in our sample, we compared our dataset with the $Kepler$ Eclipsing Binary Catalog\footnote{\url{http://$Kepler$ebs.villanova.edu/}} \citep{2016Kirk}, which comprises 2920 eclipsing/ellipsoidal binaries extracted from the complete dataset of the primary $Kepler$ mission (Q0-Q17). No binary signatures were observed in any of the sample giants. Employing a time-domain radial velocity survey could help in determining spectroscopic binary characteristics in the sample.

\subsection{Other chromospheric activity indicators}

One of the key indicators of chromospheric activity in cool stars 
is the presence of non-thermal emission reversals in the central regions of the \species{Ca}{ii}  H and K absorption lines, located at 3968.470 \r{A} and 3933.663 \r{A}, respectively. 
Chromospheric activity in cool stars is commonly expressed using the dimensionless S-index;
$$S_{\rm Ca II} = \frac{F_H + F_K}{F_B + F_R}$$ where $F_H, F_K, F_B, F_R$ are the integrated fluxes in the \species{Ca}{ii}  H, K lines over a triangular bandpass ($\Delta\lambda_{\rm HK}$) of FWHM 1.09 \r{A} and the blue and red rectangular pseudocontinuum regions of width ($\Delta\lambda_{\rm BR}$) 20 \r{A} centred around 3901.070 \r{A} and 4001.070 \r{A} respectively.

\begin{figure*}
\centering
\includegraphics[width=\textwidth]{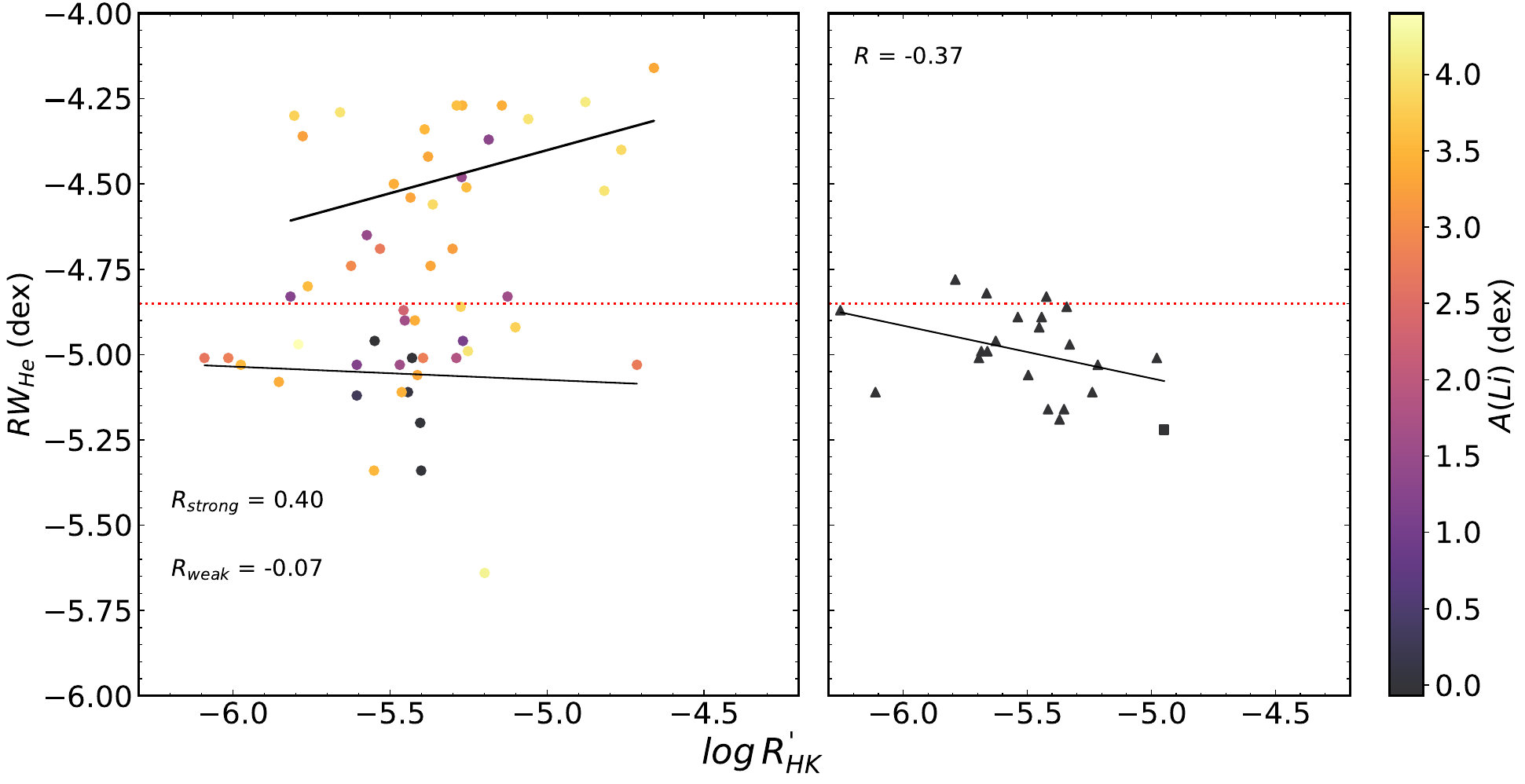}
\caption{Relationship between RW$_{He}$ and $\log R'_{\text{HK}}$ for RC (left panel) and RGB (right panel) stars. Points are color-coded based on their A(Li) values, with a square marking the position of a single subgiant. The red vertical line differentiates stars with strong and weak He I 10830 \r{A} absorption lines \citep{2022Sneden}. Solid black lines represent correlation trends within the data.
\label{fig:RWvsRhk}}
\end{figure*}
To account for the photospheric contribution to chromospheric emission and the temperature dependence of B and R fluxes a modified index is devised \citep{1984Noyes}, which is expressed as:
$$R'_{\text{HK}} = R_{\rm HK} - R_{phot}$$
where $R_{\rm HK}$ = 1.34 $\times$ 10$^{-4}$ C$_{\rm cf} \times \rm S_{MW}$, and $R_{phot}$ accounts for photospheric correction. The coefficient C$_{\rm cf}$ is dependent on (B-V) color and converts the S-index to $R_{\rm HK}$, adjusting for temperature related variations in B and R band fluxes. For evolved stars it was defined by \citet{1984Rutten}:
$$\resizebox{\columnwidth}{!}{$\log C_{cf} = -0.066(B-V)^3-0.25(B-V)^2-0.49(B-V)+ 0.45 $}$$
The photospheric contribution was given by \citet{1984Noyes}:
$$\log R_{phot} = -4.898 +1.918(B-V)^2 -2.893(B-V)^3$$
$S_{\rm Ca II}$ is converted to the Mount Wilson S-index $S_{\rm MW}$ defined by \citet{1978Vaughan}:
$$S_{\rm MW} = \alpha \times 8 \times \frac{\Delta\lambda_{\rm HK}}{\Delta\lambda_{\rm BR}}\times S_{\rm Ca II}$$
The factor of 8 is due to the design of the original Mt. Wilson spectrophotometer, which utilized a quickly rotating slit mask, leading to the H and K channels being exposed for eight times the duration of the reference pseudocontinuua channels. and $\alpha$ = 1.8 was adapted from \citet{2007Hall}. 

We calculated $S_{\rm Ca II}$ for 76 of the 84 stars that satisfied the selection criteria using ACTIN\footnote{\url{https://github.com/gomesdasilva/ACTIN2}} \citep{2021Gomes} which include spectra with S/N$_g$ $>$ 10 in the g band and fewer than 1\% negative flux values in the Ca II H and K line bandwidths.
To evaluate the accuracy of these measurements, a comparison was made with the dataset provided by \citet{2022Gehan}, which included 37 stars common to both samples. The analysis yielded a mean difference of 0.07 and a standard deviation of 0.003 between the two datasets. $S_{\rm Ca II}$ was converted to $\log R'_{\text{HK}}$ using the above relations. 

To check for a potential relationship between these two chromospheric activity indicators, we plotted RW$_{He}$ against $\log R'_{\text{HK}}$
for the sample stars in Fig \ref{fig:RWvsRhk} along with level of A(Li) in stars. 
For the RC stars (left panel), we observe a moderate positive correlation (R = 0.40) within the helium-strong (RW$_{He}$ $\geq$ -4.85) group suggesting 
overall enhanced chromospheric activity resulting in increased strengths in both the \species{He}{i} and \species{Ca}{ii} lines. In a recent study of the open cluster Stock 2 \citep{2024Jian}, a positive correlation was also found between RW$_{He}$ and $\log R'_{\text{HK}}$ for the RC population in which they observed a much tighter correlation (R=0.89).  It might be attributed to the smaller sample size, as only 9 giants were studied compared to our 76 giants. However, in the helium-weak RC group, we find a weak negative correlation (R = -0.11). This could indicate that they have reached a state of stability following previous shocks. Nevertheless, some level of basal chromospheric activity persists, which is reflected in the observed $\log R'_{\text{HK}}$ values.  It is significant to note  that 
relatively more Li-rich giants contribute to the 
positive corelation. No well defined corelation
among weaker He RC giants probably mean that strengths of \species{Ca}{ii} and \species{He}{i} evolve differentially post the enhanced activity due to the He-flash. In case of RGB stars, we observe a moderately strong negative correlation (R = -0.4). Since the He flash has not yet occurred in these stars, their chromospheric dynamics may be more stable and primarily influenced by long-term activity drivers. These stable conditions likely suppress any significant enhancement in \species{Ca}{ii}  emissions, which could explain the observed negative correlation.

\section{Discussion}\label{sec:discussion}
\begin{figure*}
\centering
\includegraphics[width=\textwidth]{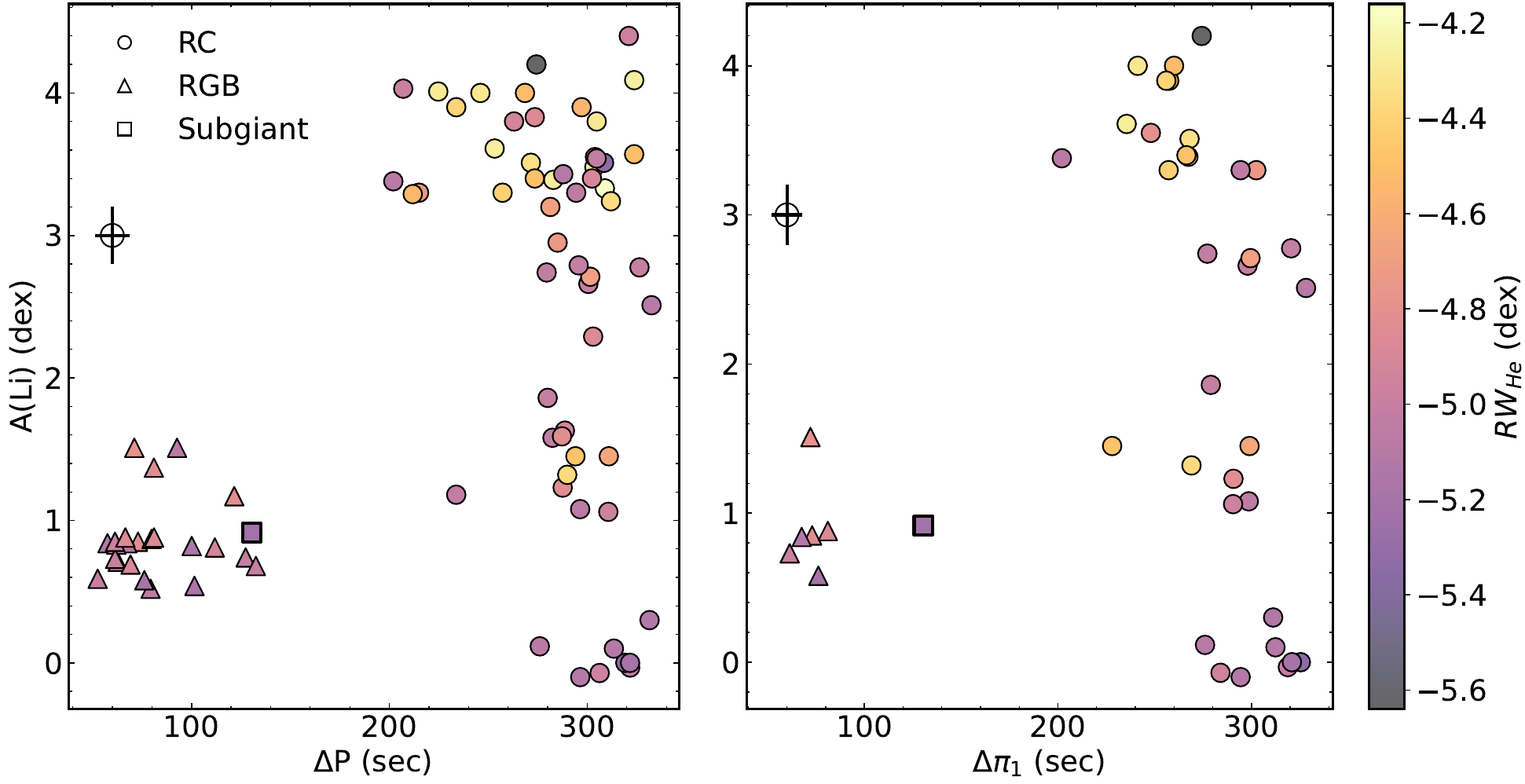}
\caption{Comparison of A(Li) against $\Delta$P (left panel) and A(Li) against $\Delta \Pi_1$ (right panel). $\Delta$P is measured for all stars in this study, while 
$\Delta \Pi_1$ which solely reflects core information, is more sensitive to the time evolution of events following the He flash and is available for 41 stars from literature. Points in both panels are color-coded according to RW$_{He}$ values}
\label{fig:PvsLivsRW}
\end{figure*}
Recently, a similar study searching for correlations between A(Li) and chromospheric \species{He}{i}  10830\AA\ was conducted on a large sample of giants \citep{2022Sneden}. This study provided a broader hint that high A(Li) and the strength of chromospheric He line are correlated; the Li-rich ( A(Li $>$ 1.5~dex)  giants are more likely to have stronger  He I strength compared to Li-poor (A (Li) $\leq$ 1.5 dex) giants.
The key difference between the current  and the previous study is that in the current study we have evolutionary phase information for the sample giants. Though the sample in the present study is relatively smaller it is better constrained in terms of the evolutionary phase. 
The data for RC and RGB giants are shown in a plot of A(Li) versus RW$_{He}$ in Figure \ref{fig:LivsHe}. 
From the figure we make the following observations;
\begin{enumerate}
\item None of the RC$_{\rm LN}$ or RGB$_{\rm LN}$ giants (see section \ref{sec:HevsLi}) are in the strong \species{He} {i} regime i.e  RW$_{He}$ $>$ 4.80 dex as defined in \citet{2022Sneden}. 
\item Of the 18 RC$_{\rm LR}$ giants 9 are He weak, 5 are He strong and four are on the uncertainty band. Among RGB$_{\rm LR}$ giants, none seems to be clearly He-strong.    
\item Majority (20 out of 29) of SLR RC Giants show strong He line strength. Four are on the broader vertical band. Remaining ones are He-Weak.  
\end{enumerate}
The key result is the clear absence of high Li abundance and strong \species{He} {i} profiles on the RGB while both features are prominent among RC giants. 
The correlation between them suggests that both strong \species{He}{i} lines and high Li in RC giants may have a  common origin, most probably the He-flash. However, it is not clear why the RGB giants have relatively weaker He $\lambda$10830 line compared to the RCs. Is it due to the He-flash in RC stars triggering higher chromospheric activity, leading to stronger He-lines, or is it because the interior He-rich material, a by-product of the H-burning shell, is dredged up to the photosphere and then the chromosphere through some flash-induced mixing event?

The He $\lambda$10830 \r{A} line strength has been studied extensively in evolved red giants \citep{1986Obrien,2004Smith,2011Dupree}. However, the impact of He-flash on this line has not been explored. The core helium flash in red giants triggers a complex series of events that can affect the chromosphere. The helium flash generates a thermal pulse, causing a rapid increase in temperature in the core. The increased thermal pressure expands the outer layers of the star rapidly. The outward-moving material from the helium flash can generate shock waves as it interacts with the layers above. These shock waves can propagate through the stellar atmosphere, disrupting the chromosphere and inducing transient dynamic disturbances. The thermal pulse, expansion and shock waves may be collectively contributing to enhanced He $\lambda$10830 absorption.
In summary, this study shows that the main He-flash and the following sub-flashes may hold the key for the enhanced He $\lambda$10830 \r{A} line among RC giants.  

In case of RGB giants, the absence of He-strong giants may be due to lesser chromospheric activity and also cooler temperatures compared to the stars post-He flash. We further discuss below the evolution of chromospheric He and the Li abundance in stars post He-flash in light of the results obtained from the asteroseismic analysis.

\subsection{Li abundance - Chromospheric He I strength correlation and their evolution post He-flash}
Apart from clear separation between giants before and after the He-flash in terms of their \species{He}{i} line strengths, we could also notice from Figure \ref{fig:LivsHe}, a steady decrease in the number of He-strong giants with decreasing Li abundances. It is also true for RGB giants as well which have mostly weak absorption profiles. This is very important to note as the SLR giants are young RCs i.e these have very recently undergone He-flash and the LN giants are old RCs (see \citet{2021Singh}). If the He-flash is the event that is driving both high Li abundance and chromospheric \species{He}{i} line strength then both these properties must be evolving with time  post He-flash. This means one would expect depleted Li abundance and lesser chromospheric activity among old RC giants. Results in Figure \ref{fig:LivsHe} provide an evidence that chromospheric He I line strength and the Li abundances are related and evolving with time. The strength of the chromospheric \species{He}{i} line appears to be linked to chromospheric activity rather than an increased He abundance. The subtle correlation between the \species{Ca}{ii} H and K indices and RW$_{He}$ among RC$_{\rm LR}$ giants provides further support for this hypothesis, suggesting that the enhanced \species{He}{i} line strengths are likely a result of heightened chromospheric activity induced by the He-flash.  

To further understand the temporal evolution of Li along with the chromospheric activity we have shown the relation between $\Delta$P, A(Li) and $RW_{He}$ in Figure \ref{fig:PvsLivsRW}. $\Delta$P is known to trace the evolution of giants' core from RGB to the RC. As shown in Figure \ref{fig:PvsLivsRW} the RGB giants (open circles) are clearly separated in the A(Li) Vs $\Delta$P plot with less A(Li) and weak He I line.  Although the A(Li) vs. $\Delta$P relation is not well defined, we observe that, on an average, younger RC stars (with relatively smaller values of $\Delta$P) tend to have more SLRs with strong He I lines, compared to older RC giants (with $\Delta$P $\sim$ 320 sec), which are mostly Li-normal with weak \species{He}{i} lines. The relation is better noticeable in a plot of A(Li) versus $\Delta\Pi_{1}$ plot.  The asymptotic period spacing ($\Delta\Pi_{1}$) of the dipole g-mode is understood to be a better representative asteroseismic parameter linked to the core evolution. However, this is only computed for giants which have much better quality data with long cadence. The younger RCs with an average $\Delta\Pi_{1}$ $\sim$ 260s are found to be more likely to be SLR with strong \species{He}{i} lines compared to older RCs with an average value of $\Delta\Pi_{1}$ $\sim$ 300s (see \citet{2021Singh}).

\section{Conclusions}\label{sec:conclusion}
In this study, we analyzed asteroseismic data, chromospheric \species{He}{i} 10830\r{A} line strengths, and photospheric lithium abundances for 84 giants in the $Kepler$ field to investigate the origin of high Li in RC giants. Our findings reveal a clear distinction between RGB and RC giants in both Li abundance and chromospheric activity. RGB giants exhibit subdued chromospheric activity, weaker \species{He}{i} lines, and lower Li, while RC giants are characterized by strong \species{He}{i} lines and high Li abundance. Notably, we observe a decline in the number of He-strong giants with decreasing Li abundance, consistent with the transient nature of high Li among RC giants.

Our results suggest that He-strong and Li-rich giants are likely younger RC stars, whereas Li-normal and He-weak RC giants are older. The presence of a few SLR giants with weak \species{He}{i} lines indicates that these properties may evolve on different timescales, reflecting variations in the impact of the He-flash. Furthermore, we find stronger \species{Ca}{ii} H \& K emission indices among SLR giants, supporting the hypothesis that this transient enhanced chromospheric activity due to the He-flash contributes to both \species{He}{i} and \species{Ca}{ii} features.

Looking ahead, it remains unclear whether the strength of \species{He}{i} line is more influenced by local chromospheric conditions like density and temperature, or by transient disturbances caused by the core He-flash. Modeling the chromosphere would help clarify this, and in doing so, we can translate our measured EWs into chromospheric He abundances. Additionally, as suggested by \citet{2014Hema}, their method using MgH bands in optical spectra (for cool stars without photospheric He lines) could help establish relations between photospheric and chromospheric He abundances in connection with the He flash. This would require higher resolution optical spectra. Asteroseismic data can also be used to study acoustic glitches from the He ionization zone. By calibrating these glitches against models of known He  abundance \citep{2014Verma}, we could determine photospheric He abundances, offering an alternative to spectroscopic methods.

\section*{acknowledgments}
These results are derived from data collected using the Habitable-zone Planet Finder Spectrograph mounted on the Hobby-Eberly Telescope. The authors extend their gratitude to the Telescope Operators at HET for the precise execution of observations using HPF. The Hobby–Eberly Telescope is a collaborative effort between the University of Texas at Austin, Pennsylvania State University, Ludwig-Maximilians-Universität München, and Georg-August Universität Göttingen. The HET is named in recognition of its principal benefactors, William P. Hobby and Robert E. Eberly. The collaboration acknowledges the assistance and resources provided by the Texas Advanced Computing Center. The authors also wish to acknowledge funding from NSF grants AST-1616040 (CS), AST-1908892 (GNM), and TÜBITAK project No. 112T929 (MA).

This work makes use of publicly available data from LAMOST DR10 v2.0 (released September 2024). The Guoshoujing Telescope (LAMOST), a National Major Scientific Project, was built by the Chinese Academy of Sciences. Funding for LAMOST has been provided by the National Development and Reform Commission, and the telescope is operated and managed by the National Astronomical Observatories, Chinese Academy of Sciences.

We gratefully acknowledge the entire $Kepler$ and $TESS$ teams, as well as all those involved in the missions, for their invaluable contributions. Funding for the Kepler Mission is provided by NASA's Science Mission Directorate. All of the data presented here were obtained from the Mikulski Archive for Space Telescopes (MAST).
\facilities{HET (Habitable-zone Planet Finder Spectrograph),LAMOST,MAST}
\software{linemake (\citet{2021Placco},\url{https://github.com/vmplacco/linemake}),  
IRAF \citep{1986Tody,1993Tody}, 
SPECTRE \citep{1987Fitzpatrick,2012Sneden}, 
Goldilocks (\url{https://github.com/grzeimann/Goldilocks_Documentation}), pyMOOGi (\url{https://github.com/madamow/pymoogi}), Lightkurve \citep{2018Lightkurve}, pySYD \citep{2022Chontos}, ACTIN 2 \citep{2018Gomes,2021Gomes}}

\begin{deluxetable*}{ccccccccccccccccc}
\rotate
\tabletypesize{\scriptsize} 
\tablewidth{\textwidth} 
\tablecaption{Derived and Adopted Parameters of the \textit{$Kepler$} Sample \label{tab:1}}
\tablehead{\colhead{KIC} & \colhead{T$_{\rm eff}$} & \colhead{log \textit{g}} & \colhead{[Fe/H]} & \colhead{A(Li)} & \colhead{Source} & \colhead{EW$_{\rm He}$} & \colhead{RW$_{\rm He}$} & \colhead{RUWE} & \colhead{$\nu_{\rm max}$} & \colhead{$\Delta \nu$} & \colhead{$\Delta$P} & \colhead{Evol stage}\tablenotemark{a} & \colhead{Mass} & \colhead{$\log \left (\frac{L}{L_{\odot}}\right )$} & 
\colhead{(B-V)} & \colhead{$\log R'_{\rm HK}$} \\
\colhead{} & \colhead{(K)} & \colhead{(dex)} &\colhead{(dex)} &\colhead{(dex)} &\colhead{} &\colhead{(m\AA)} &\colhead{(dex)} &\colhead{(dex)} &\colhead{($\mu$Hz)} & \colhead{($\mu$Hz)} &\colhead{(s)} &\colhead{} &\colhead{(M$_{\odot}$)} & \colhead{(L$_{\odot}$)} & \colhead{(dex)} &\colhead{(dex)}}
\startdata
1726211 & 4965±95 & 2.24±0.06 & -0.61±0.03 & -0.03±0.17 & 3 & 105±15 & -5.01±0.06 & 0.90 & 30.40±0.63 & 3.72±0.05 & 321.66±6.50 & 1 & 1.44±0.35 & 1.82 & 0.98±0.19 & -5.43±0.50 \\ 
2305930 & 4879±25 & 2.47±0.06 & -0.39±0.03 & 3.90±0.03 & 4 & 300±10 & -4.56±0.01 & 0.91 & 27.92±0.94 & 3.77±0.10 & 297.12±10.00 & 1 & 0.74±0.13 & 1.60 & 1.18±0.20 & -5.36±0.29 \\ 
2449858 & 4840±30 & 2.50±0.10 & -0.15±0.00 & 3.30±0.28 & 1 & 195±10 & -4.74±0.02 & 0.95 & 26.76±0.45 & 3.46±0.05 & 215.00±6.60 & 1 & 1.23±0.12 & 1.80 & 1.18±0.02 & -5.37±0.03 \\ 
2714397 & 5003±100 & 2.44±0.01 & -0.62±0.15 & 0.00±0.19 & 3 & 50±10 & -5.34±0.09 & 1.02 & 33.08±0.52 & 4.18±0.04 & 319.22±3.46 & 1 & 1.10±0.06 & 1.71 & 1.04±0.15 & -5.40±0.28 \\ 
3748691 & 4954±100 & 2.50±0.01 & 0.07±0.15 & 0.00±0.26 & 3 & 69±10 & -5.20±0.06 & 0.84 & 38.71±0.79 & 4.24±0.05 & 321.61±3.97 & 1 & 1.63±0.07 & 1.76 & 0.81±0.36 & -5.40±1.84 \\ 
3751167 & 4914±80 & 2.33±0.03 & -0.76±0.15 & 4.00±0.57 & 1 & 330±15 & -4.52±0.02 & 1.02 & 26.14±1.59 & 3.59±0.12 & 268.52±17.60 & 1 & 0.95±0.22 & 1.80 & 1.23±0.20 & -4.82±0.28 \\ 
3858850 & 4375±14 & 2.23±0.03 & 0.29±0.02 & 2.95±0.09 & 4 & 195±10 & -4.74±0.02 & 0.96 & 25.92±0.69 & 3.48±0.07 & 285.00±15.70 & 1 & 0.95±0.11 & 1.63 & 1.32±0.14 & -5.62±0.25 \\ 
4044238 & 4702±164 & 2.44±0.01 & 0.00±0.30 & 1.08±0.35 & 3 & 95±8 & -5.06±0.04 & 1.05 & 34.55±0.75 & 4.01±0.05 & 296.40±3.06 & 1 & 1.33±0.21 & 1.61 & ~ & ~ \\ 
4161005 & 4897±40 & 2.35±0.10 & -0.52±0.00 & 3.30±0.36 & 1 & 415±15 & -4.42±0.02 & 1.02 & 29.10±0.96 & 3.90±0.12 & 257.20±1.89 & 1 & 1.03±0.18 & 1.67 & 1.09±0.03 & -5.38±0.06 \\ 
4446405 & 4846±100 & 2.69±0.01 & -0.13±0.15 & 1.37±0.19 & 5 & 165±10 & -4.82±0.03 & 0.89 & 59.96±0.65 & 5.75±0.02 & 81.04±3.90 & 0 & 1.58±0.09 & 1.55 & 1.33±0.51 & -5.66±0.76 \\
\enddata
\tablerefs{(1) \citet{2019Singh} (2) \citet{2021Singh} (3) \citet{2017Takeda} (4) \citet{2021Yan} (5) Present work}
\tablenotetext{a}{0-RGB, 1-RC, 2-subgiant}
\tablecomments{Only a portion of this table is shown here to demonstrate its form and content. A machine-readable version of the full table is available.}
\end{deluxetable*}
\clearpage
\bibliography{draft1}{}
\bibliographystyle{aasjournal}
\end{document}